\documentstyle[12pt,fleqn]{article}
\begin{document}
\vspace*{2cm}
\noindent
{\large Kondo resonance in the case of strong
Coulomb screening }
\vspace{0.2in} \\
V. V.  Ponomarenko
\vspace{0.2in} \\
RIKEN, Frontier Research System,
2-1 Hirosawa, Wako-Shi, Saitama 351-01, Japan  \\
Permanent address: A.F.Ioffe Physical Technical Institute, 194021,
St. Petersburg, Russia\\
\vspace{0.3in}

The effect of Coulomb screening on the magnetic impurity behavior is analysed.
Two types of the behavior corresponding to
either integer or
fractional occupation numbers  of the low lying magnetic level are described.
The
features in the dependence of the resistivity on the magnetic field
specific for these regimes are predicted.

\vspace{0.2in}
\noindent
{\bf 1. MODEL}
\vspace{0.2in}

   There is a mechanism of the Coulomb screening of charging of a localized
electronic  state located inside a conductor, which  was first
discussed by G. Mahan, P. Nozieres and C. De Dominicis, and by K. D. Schotte
and U. Schotte $[1]$. They demonstrated that the rate of a one-electron jump
$1/\Gamma, \ (\hbar=1)$ undergoes exponential renormalization due to both the
reconstruction of the Fermi seas and the exchange effect. It was supposed
later [2] that the same mechanism pertains to the magnetic impurity behavior.
I will examine the effect on the properties of
a magnetic impurity
by calculating the zero temperature impurity resistivity $\rho$ as a function
of magnetic field $H$ in two regimes of the impurity behavior.
Suggestion of the existence of the two regimes is based on the Bethe Ansatz
solution of the model of the resonant level in the spinless case [3]. It
confirmed that for either a relatively weak Coulomb interaction or
a wide enough
conduction band the renorm-group approach extending  the perturbation
expansion in $\Gamma $ brings correct results out.
At the same time, in the case
of a strong exchange effect ensuing from the strong Coulomb interaction the
solution becomes non-perturbational in  $\Gamma $ for a level lying low
inside the conduction  band and predicts its fractional occupation.
\par
 Following $[2]$, the impurity with $(2j+1)$ degenerate
localized magnetic state $d_m, m=-j,...j$
could be specified by
a generalized Anderson model with the Hamiltonian
\begin{eqnarray}
{\it H }  =  \int_{-D}^{D} d \epsilon \left( \sum_{\gamma} \epsilon
c^+_{ \gamma}( \epsilon) c_{ \gamma} ( \epsilon)+
\sqrt{ \Gamma/ \pi} \sum_m [c^+_{m}(\epsilon)d_m+
d^+_m c_{m}(\epsilon)] \right) + \nonumber \\
\sum_m \left( \sum_{\gamma} \frac{U_{\gamma}}{2\pi}
\int \int d\epsilon d\epsilon'
c^+_{\gamma}(\epsilon)c_{\gamma}(\epsilon')+\epsilon_D+mH \right)d^+_md_m
\end{eqnarray}
where the energy of the state $\epsilon_D \ll -\Gamma, -|H|$ is much lower
than the chemical potential $\mu=0$, and
the full occupation number $\sum \hat{n}_m, \hat{n}_m=d^+_m d_m$
is limited to zero or one because
of a strong intrasite interaction. The 1D channel states of the
continuous spectrum
$ c_{\gamma}(\epsilon), |\epsilon|<D $ are tunneling for $ \gamma=m $,
if they are
of the same symmetry as the appropriate orbital $ d_m $. Otherwise, for
$ \gamma=i_{sc} $, they only screen the charging of the impurity
acquiring the phase shifts the sum of which is $ \varphi_{sc}= -
\sum_i arctan(U_i/2) $. The phase shift due to
scattering on the constant charge
of the impurity
in the tunneling channels is $ \varphi= -arctan(U/2), U_m=U $.

Accounting for the screening mechanism in Eq. (1) results in additional
phase shifts important for the resistivity calculation and in the modification
of the tunneling processes.
The contribution to the resistivity is expressed by the phase shifts
of the tunneling channels $ \delta_m$ as
$ \rho=(2j+1)\rho_0/(\sum (sin\delta_m(H))^{-2}) $,
where $\rho_0$ is a dimensional
constant independent of $H$.
To calculate these phase shifts one can generalize
Friedel's Sum Rule to our case $ \delta_m=\pi n_m-(\pi + \varphi_{sc})
\sum_{m'} n_{m'}/(2j+1) $, where the condition of the charge neutrality
$\varphi_{sc}+(2j+1)\varphi=-\pi$ is accounted for.
 Below I will present results for $j=1/2$, though both the
qualitative predictions and  the considerations
remain true in the general case.

\vspace{0.2in}
\noindent
{\bf 2. INTERGER OCCUPATION REGIME}
\vspace{0.2in}

In the case of a relatively weak exchange effect ( following [3]
I suppose $ D \gg \Gamma U $ ) the modification of the tunneling
process mainly reduces to an increase of the tunneling half-width by
a factor of about $(D/\Gamma)^{\beta}$, where $\beta$ is expressed by
the phase shifts $[1]$.
For the low lying level the full occupation number is close to
one and
\begin{equation}
\rho=\frac{2 \rho_0}{[sin(\varphi_{sc}/2+\pi \Delta n/2)]^{-2}+
[sin(\varphi_{sc}/2-\pi \Delta n/2)]^{-2} }, \ \ \ \Delta n(H)=
n_{1/2}-n_{-1/2} \label{2}
\end{equation}
The magnetization $\Delta n $ is a smooth antisymmetric function of $H$.
It varies from -1 to 1 when $H$ are passing from $-H_K$ to $H_K$
($H_K$ is the Kondo energy). Then Eq.(\ref{2}) means
that two minima of $\rho(H)$ dependence exist (fig.1). They
are located at
$\pm H_x$, where $ \Delta n(H_x)=\varphi_{sc}/ \pi $.
$H_x$ varies from $\infty$
for $\varphi=0$ ( usual Anderson model) to $H_x=0$ for $\varphi_{sc}=0$.

\begin{figure}
  \vspace{4.5cm}
  \caption{ Qualitative dependence of the impurity resistivity $\rho$
  on magnetic field $H$ for $\varphi_{sc}=-\pi$ (dashed line),
   $\varphi_{sc}=0$ (dotted line), $\varphi_{sc}=-2\pi/3$ (solid line) in the
   integer occupation regime }
\end{figure}
In fact, $\delta_m$ is zero at $H=0$ for any $j$ if $\varphi_{sc}=0$.
This proves
that $\rho$ has a minimum at $H=0$, instead of a maximum as for
$\varphi_{sc}=-\pi$.

\vspace{0.2in}
\noindent
{\bf 3. FRACTIONAL OCCUPATION REGIME}
\vspace{0.2in}

If $ \varphi $ is close to $ -\pi/2 \ \
( \varphi_{sc} \approx 0)$
and $D \ll \Gamma U $ the behavior of the impurity changes drasticly.
The Pauli's principle does not prevent a localized electron from
tunneling anymore
if all wave functions of the filled part of the conduction band
decrease to zero near the location of the localized impurity state
due to Coulomb
repulsion. Owing to the occurence of a special symmetry $[3]$ the Hamiltonian
Eq. (2) allows a Bethe Ansatz solution. Indeed, the artificial
two-particle $S$-matrices neccessary to construct the collective wave-function
$[4]$
take the form
\begin{equation}
\hat{S}=- \frac{\epsilon_1- \epsilon_2+2i \Gamma \hat{P}}
{\epsilon_1- \epsilon_2-2i \Gamma }
\end{equation}
where $\epsilon_1,\epsilon_2$ are the rapidities of the scattering particles
and $\hat{P}$ permutes their spins. These $S$-matrices are unitary and
satisfy Yang-Baxter equations. It allows to construct the solution following
a standard
scheme [5]. In this scheme the system is considered to be on the
ring of the length $L$,
and the periodic boundary conditions are imposed. This leads to the equations
of the spectrum of the system.
\begin{eqnarray}
\lefteqn{
e^{i\epsilon_l L}=(-1)^N \frac{\epsilon_l+i\Gamma}{\epsilon_l-i\Gamma}
\prod_{j=1}^N \frac{\epsilon_l-\epsilon_j-i2\Gamma}
{\epsilon_l-\epsilon_j+i2\Gamma}
\prod_{k=1}^m \frac{\epsilon_l-\xi_k+i\Gamma}{\epsilon_l+\xi_k-i\Gamma}
 }\\
\lefteqn{
\prod_{j=1}^N \frac{\xi_l-\epsilon_j+i\Gamma}
{\xi_l-\epsilon_j-i\Gamma}=-\prod_{k=1}^m
\frac{\xi_l-\xi_k+i2\Gamma}{\xi_l+\xi_k-i2\Gamma} \nonumber
}
\end{eqnarray}
where $\epsilon_l$ and $\xi_l$ are the rapidities of the holons and spinons.
Their numbers correspond to the full number of electrons and the number of
electrons with turned over spin (spin down), respectively. Minimizing of the
energy of the system $ \sum \epsilon_j $ under the
condition $ Re \epsilon_j > -D $
one can find [6] that all rapidities are real in the ground state. Finally,
these equations
could be reduced to the equations on the densities of charge
$r_1(\epsilon_l)=1/(L|\epsilon_{l+1}-\epsilon_l|)$ and
spin $r_2(\xi_k)=1/(L|\xi_{k+1}-\xi_k|)$ rapidities
\begin{eqnarray}
\lefteqn{
\frac{1}{2\pi}+\frac{1}{L\pi}\Phi_2(\epsilon-\varepsilon_D)=
r_1(\epsilon)+\frac{1}{\pi} \int_{-D}^{\mu_1} d \epsilon'
\Phi_1(\epsilon-\epsilon')r_1(\epsilon')-
\frac{1}{\pi} \int_{-\infty}^{\mu_2} d \xi\Phi_2(\epsilon-\xi)r_2(\xi)
 \label{5}
 }\\
\lefteqn{
\frac{1}{\pi} \int_{-D}^{\mu_1}d \epsilon\Phi_2(\xi-\epsilon)r_1(\epsilon)
=r_2(\xi)+
\frac{1}{\pi} \int_{-\infty}^{\mu_2}d \xi'\Phi_1(\xi-\xi')r_2(\xi')
 \label{6}
 } \\
\lefteqn{
\Phi_n(\epsilon)=\frac{n}{2\Gamma} \frac{1}{1+(n \epsilon/2 \Gamma)^2}
\nonumber }
\end{eqnarray}
Here $\varepsilon_D$ differs from $ \epsilon_D $ by a constant due to
ambiguity of the logarithm. Both densities consist of the volume and
impurity parts $ r_a=r_{a,h}+r_{a,imp} $, which are produced by
$1/(2\pi)$ and $1/(L\pi) \Phi_2(\epsilon-\varepsilon_D)$ on the left-hand side
of (\ref{5}), respectively. $r_{a,h}$ are connected with the electron density
$n_h=N/L=n_{h,1/2}+n_{h,-1/2}$ by
\begin{equation}
n_h=\int_{-D}^{\mu_1} d \epsilon r_{1,h}(\epsilon), \ \ \
n_{h,-1/2}=\int_{-\infty}^{\mu_2} d \xi r_{2,h}(\xi)
\end{equation}
This should be used to define $\mu_1$ and $\mu_2$.
Then the impurity occupation numbers
are calculated as
\begin{equation}
n_{1/2}+n_{-1/2}=\int_{-D}^{\mu_1} d \epsilon r_{1,imp}(\epsilon), \ \ \
n_{-1/2}=\int_{-\infty}^{\mu_2} d \xi r_{2,imp}(\xi)
\end{equation}
\par
To describe the behavior of the system in this regime I have
solved Eq. (\ref{5}, \ref{6}) in the limit $\Gamma \rightarrow 0$ when all
$\Phi_n(\epsilon)$ transform into a delta-function. The solution shows that:\\
$r_2= r_1/2, r_1(\epsilon)=1/(3\pi)+2/3 \delta(\epsilon -\varepsilon_D)$
for the spin-degenerate filled states $ \epsilon < \mu_2 $ ; \\
$r_2=r_1(\epsilon)=1/(4\pi)+1/2 \delta(\epsilon -\varepsilon_D)$
for the spin-nondegenerate filled states $ \mu_2<\epsilon < \mu_1 $ ; \\
$r_2=0, r_1(\epsilon)=1/(2\pi)+\delta(\epsilon -\varepsilon_D)$
for the empty states $\mu_1<\epsilon$.\\
These results mean that the level remains spin-degenerate with the fractional
occupation 1/3 for each spin component unless $\varepsilon_D$ becomes
greater than
$\mu_2 $. It is spin-nondegenerate with the fractional occupation number equal
to 1/2 if $\mu_2<\varepsilon_D<\mu_1 $. The difference $\mu_1-\mu_2 = 2H$
shows that all energy parameters of the level undergo
renormalization by a factor
equal to the fractional occupation number. Returning to the impurity
resistivity
one can use its expression in terms of the occupation numbers to
conclude that there is no field dependence of the impurity
resistivity
for low lying level $\varepsilon_D \ll 0$ unless $H$ becomes more than
$|\varepsilon_D |$.

\vspace{0.2in}
\noindent
{\bf 4. CONCLUSION }
\vspace{0.2in}

It was shown that there is a special behavior of the magnetic level lying
low inside the coduction band. It occurs under the condition of
resonant Coulomb screening when all filled states of the conduction band,
deformed by the Coulomb interaction with a localized electron, become about
orthogonal to the state of the electron jumping out from the impurity level.
Such a behavior is characterized by the fractional occupation number equal to
$(2j+1)/(2j+2)$ for the $(2j+1)$ degenerate level
( $1/(2j+2)$  for each component ). The degeneracy cannot be removed unless
the magnetic field $H$ becomes greater than the energy of the level.
This behavior is quite different
from the other one typical in the case of a weak Coulomb interaction
which develops
smoothly under the switching on of the Coulomb screening of the impurity
charging. The latter corresponds to
an integer occupation number. Its degeneracy is
removed by the small magnetic field
$ H_K \approx D exp[-\varepsilon_D/\Gamma] $, what leads to a well-known
hump in the
$\rho(H)$ dependence located at $H=0$. The width of the hump becomes much
less than $H_K$ if the screening effect is essential.

\vspace{0.3in}
\noindent
{\bf REFERENCES }
\vspace{0.3in} \\
\noindent
 1. G.D. Mahan, Phys. Rev. {\bf 163}, (1967) 612;
 P. Nozieres and C. DeDominisic, Phys.

\noindent
\hspace{0.2in}
 Rev.{\bf 178}, (1969) 1097;
K. D. Schotte and U. Schotte,
Phys. Rev.{\bf 182}, (1969) 479.

\noindent
 2. F.D.M. Haldane, Phys. Rev. B {\bf 15}, (1977) 2477.

\noindent
 3. V. V. Ponomarenko, Phys. Rev. B {\bf 48}, (1993) 5265;
Phys. Rev. B {\bf 51}, (1995) 5498.

\noindent
 4. A.M. Tsvelick and P.B. Wiegmann, Adv. in Phys.{\bf 32 }, (1983) 453.

\noindent
 5. M. Gaudin, La Fonction d'Onde de Bethe (Masson, Paris, 1983).

\noindent
 6. P. Schlottman, Phys. Rev. B {\bf 41}, (1990) 4164.

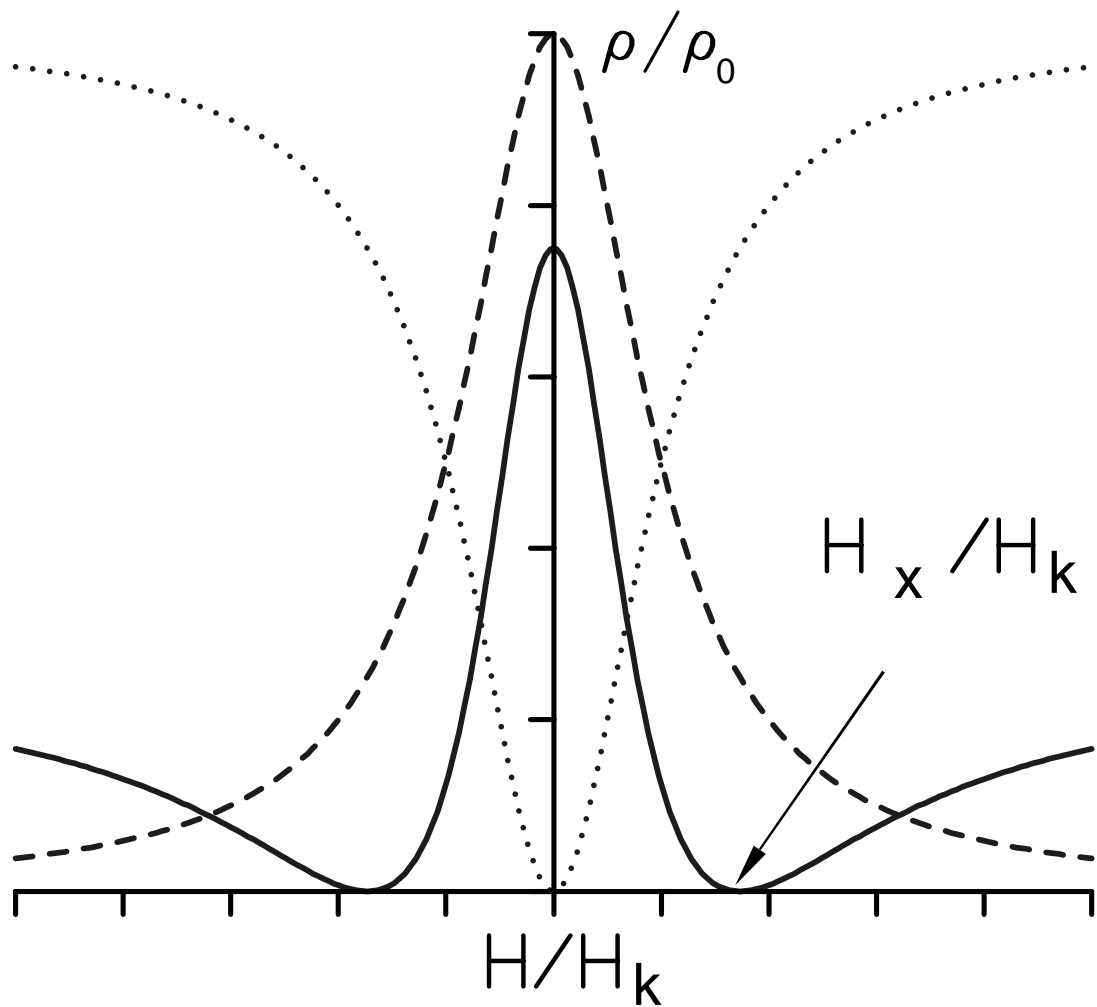
\end{document}